%% file: main.tex
\documentclass[10pt,conference]{IEEEtran}
\usepackage{hyperref}
\usepackage{siunitx}
\usepackage{dcolumn}
\IEEEoverridecommandlockouts

\usepackage{cite}
\usepackage{amsmath,amssymb,amsfonts}
\usepackage{algorithmic}
\usepackage{graphicx}
\usepackage{textcomp}
\usepackage{xcolor}
\usepackage{listings}
\usepackage{color}
\usepackage{booktabs}
\usepackage{tabularx}
\usepackage{enumitem}
\usepackage{xspace}
\definecolor{dkgreen}{rgb}{0,0.6,0}
\definecolor{gray}{rgb}{0.5,0.5,0.5}
\definecolor{mauve}{rgb}{0.58,0,0.82}
\hypersetup{colorlinks=true, unicode=true, linkcolor=[rgb]{0.10,0.05,0.67}, citecolor=[rgb]{0.10,0.05,0.67}, filecolor=[rgb]{0.10,0.05,0.67}, urlcolor=[rgb]{0.10,0.05,0.67}}
\def\BibTeX{{\rm B\kern-.05em{\sc i\kern-.025em b}\kern-.08em
    T\kern-.1667em\lower.7ex\hbox{E}\kern-.125emX}}
\newcommand{\ie}{\textit{i.e.,}\xspace}
\newcommand{\eg}{\textit{e.g.,}\xspace}
\newcommand{\etal}{\textit{et al.}\xspace}

\newcommand{\mysubsection}[1]{\vspace{1.25mm}\noindent\textit{\textbf{#1.}}}
\newcommand{\mysubsubsection}[1]{\vspace{1.25mm}\noindent\textit{#1.}}
\newcommand{\smalltt}[1]{{\small \texttt{#1}}}
    
\begin{document}

\title{Learning from Mistakes: Understanding Ad-hoc Logs through Analyzing Accidental Commits}
 \author{
 \IEEEauthorblockN{Yi-Hung Chou}
 \IEEEauthorblockA{
 \textit{University of California, Irvine}\\
 Irvine, United States \\
 yihungc1@uci.edu}
 \and
 \IEEEauthorblockN{Yiyang Min}
 \IEEEauthorblockA{
 \textit{Amazon, Toronto}\\
 Toronto, Canada \\
 rogermyy@amazon.com}
 \and
 \IEEEauthorblockN{April Yi Wang}
 \IEEEauthorblockA{
 \textit{ETH Zürich}\\
 Zürich, Switzerland \\
 april.wang@inf.ethz.ch}
 \and
 \IEEEauthorblockN{James A. Jones}
 \IEEEauthorblockA{
 \textit{University of California, Irvine}\\  Irvine, United States \\
 jajones@uci.edu}
}

\newif\ifcomment

\def\comment#1{\ifcomment\relax\else#1\fi}

% \commenttrue

\maketitle

\begin{abstract}

Developers often insert temporary ``print'' or ``log'' instructions into their code to help them better understand runtime behavior, usually when the code is not behaving as they expected. 
Despite the fact that such monitoring instructions, or ``ad-hoc logs,'' are so commonly used by developers, there is almost no existing literature that studies developers' practices in how they use them.
This paucity of knowledge of the use of these ephemeral logs may be largely due to the fact that they typically only exist in the developers' local environments and are removed before they commit their code to their revision control system.
In this work, we overcame this challenge by observing that developers occasionally mistakenly forget to remove such instructions before committing, and then they remove them shortly later.
Additionally, we further studied such developer logging practices by watching and analyzing live-streamed coding videos.
Through these empirical approaches, we presented where, how, and why developers use ad-hoc logs to better understand their code and its execution.
We collected 27 GB of accidental commits that removed 548,880 ad-hoc logs in JavaScript from GitHub Archive repositories to provide the first large-scale dataset and empirical studies on ad-hoc logging practices.
Our results revealed several illuminating findings, including a particular propensity for developers to use ad-hoc logs in asynchronous and callback functions.
Our findings provided both empirical evidence and a valuable dataset for researchers and tool developers seeking to enhance ad-hoc logging practices, and potentially deepen our understanding of developers' practices towards understanding of software's runtime behaviors.
    
\end{abstract}


\input{src/introduction}
\input{src/background}

\input{src/dataset}

\input{src/empirical_study}
\input{src/discussion}
\input{src/related_works}
% \section{Acknowledgment}

\bibliographystyle{IEEEtran}
\bibliography{references, data}

\end{document}

%% file: src/introduction.tex
\section{Introduction}
Logs, such as \textit{print} in Python or \textit{console.log} in JavaScript, are often used for the following two scenarios:

\begin{description}[leftmargin=*]
    \item[Production Logs:] Developers insert logging statements to continuously monitor the software after deployment \textbf{(\ie logs in production)}~\cite{chen_survey_2022}. For instance, they might insert a logging instruction into an exception block to record the errors that occur during production, such as \smalltt{logger.error(e.errorMessage)}. These logs are considered essential for monitoring active software in production and are necessary to continue to be kept in the codebase.\vspace{1mm}
    \item[Ad-hoc Logs:] Developers insert logging statements on their local machines for quickly understanding the runtime behaviors of the software under development \textbf{(\ie ad-hoc logs)}~\cite{layman_debugging_2013, jiang_log-it_2023}. For instance, if a crash occurs during development, a developer might hypothesize, ``I do not think the program executes this line of code.'' They would then temporarily insert a log statement at the target location to collect runtime information, verifying whether the runtime behaviors match their hypotheses, and continuing their experiments accordingly and iteratively. Such print statements are intended to be deleted before developers push their code into production or version-control system. 
\end{description}

Many software-engineering studies have primarily focused on production logs (\eg \cite{li_qualitative_2021, chen_survey_2022, gholamian_comprehensive_2022, fu_where_2014, liu_which_2021, li_where_2020}). For example, some research examines the optimal placement of production log statements (\eg \cite{fu_where_2014, li_where_2020}), while other studies analyze the content or specific variables captured in log statements (\eg \cite{liu_which_2021, li_are_2023}). Building on these empirical studies, researchers have also proposed tools that suggest the best locations and content for log messages to enhance logging practices (\eg \cite{zhu_learning_2015, fu_where_2014, mastropaolo_using_2022}). Additionally, with the large volume of log messages generated, other studies explore which logs developers should prioritize~\cite{cinque_what_2014} or aim to improve the usability of log outputs through visualizations~\cite{jiang_log-it_2023}.
\begin{figure}[t]
    \centering
    \includegraphics[width=0.85\linewidth]{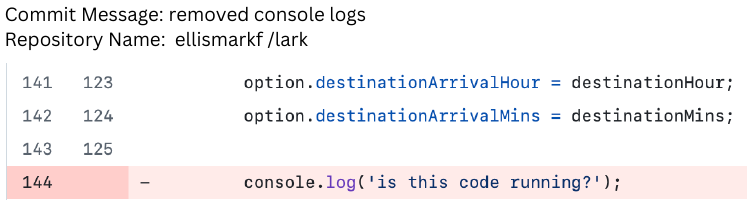}
    \caption{Example GitHub commit that removes a previous accidental commit of an ad-hoc log}
    \label{fig:exampler_logs}
\end{figure}
In contrast, to the best of our knowledge, the second type of logs---\emph{ad-hoc logs}\footnote{In this work, we use ``log'' and ``logging statement'' interchangeably to refer to a logging instruction, as in prior work~\cite{jiang_log-it_2023}.}---remains largely understudied in software-engineering literature~\cite{chen_survey_2022}. Despite their essential role in software development, particularly in debugging~\cite{alaboudi_exploratory_2019, alaboudi_what_2023, jiang_log-it_2023, beller_dichotomy_2018, siegmund_studying_2014, noauthor_studying_nodate}, the transient and local nature of ad-hoc logs has limited the research community’s ability to study and understand them.

% Fortunately, for the purposes of research, not all \textbf{ad-hoc logs} are completely hidden in developers' local environments. Like anyone, developers occasionally make mistakes when pushing code to version-control systems, ranging from simple, easily fixable errors~\cite{wen_empirical_2020} to unintentionally exposing sensitive information, such as API keys~\cite{meli_how_2019}. Our research leverages this serendipitous opportunity to address a significant research gap: understanding ad-hoc logs through mining accidental commits of them, and those that remove them, as indicated by commit messages (\eg ``remove console.log'') in version-control systems like GitHub. Figure~\ref{fig:exampler_logs} shows one example of commits and logs that we mined from the GitHub Archive.

Fortunately, for the purposes of research, not all \textbf{ad-hoc logs} are completely hidden in developers' local environments. Like anyone, developers occasionally make mistakes when pushing code to version-control systems, such as GitHub, ranging from simple, easily fixable errors~\cite{wen_empirical_2020} to exposing sensitive information, such as API keys~\cite{meli_how_2019}. Our research leverages this serendipitous opportunity to address a significant research gap: inferring developer challenges in understanding runtime behavior of code by analyzing large-scale empirical evidence of ad-hoc logs.
% \yihung{
% \begin{quote}
%     Approximate the comprehension of dynamic aspect of software by analyzing large-scaled empirical evidences of ad-hoc logs.
% \end{quote}
% }

This research enables such understanding through mining accidental commits that remove them, as indicated by commit messages (\eg ``remove console.log'') in version-control systems like GitHub. Figure~\ref{fig:exampler_logs} shows one example of commits and logs that we mined from the GitHub Archive.

For deepening our understanding of the use of these logs, we also qualitatively analyzed the 36 hours of developers' live-stream videos collected by Alaboudi \etal~\cite{alaboudi_exploratory_2019}. This further analysis allows us to see all of the temporary placement and use of their ad-hoc logs, instead of just the ones that were mistakenly committed to the version-control system. It also allows us to better understand the situations and comprehension tasks in which the developer needed to use them. Whereas the GitHub data gives us breadth and scalability, the live-stream data gives us depth and qualitative insight.

The results of our studies revealed a number of interesting findings.
For example, we found a common logging behavior in which developers made efforts in structuring the console outputs so that they could pinpoint and compare the program states or executions on the fly.

Additionally, we found that certain types of code locations that developers used ad-hoc logs were overrepresented when looking at the proportion of those types of locations in the overall dataset.
For example, developers placed these print statements into asynchronous and callback functions at a greater rate than the overall rate of those types of functions in the dataset.
Apparently, such functions required logging for developers to understand their execution. 
Moreover, developers used these logs to understand more complex-logic functions, more often than those of lesser complexity.

Our study makes the following contributions: 
(1) We conducted an empirical study for understanding  ad-hoc logs, guided by research questions informed by prior literature and qualitative findings. 
(2) We analyzed developers' live-coding sessions~\cite{alaboudi_exploratory_2019, alaboudi_what_2023} 
to gain insights into their ad-hoc logging behaviors, which provides qualitative context for understanding the motivations and ad-hoc log-authoring practices. 
(3) We introduced a novel mining method that captures data previously considered inaccessible or hidden in local environments, such as ad-hoc logs, by leveraging developer mistakes.
(4) We collected and published 27 GB of JavaScript commits from GitHub Archive~\cite{github-archive}, and Google Cloud BigQuery~\cite{bigquery}, containing 548,880 removed logs, along with corresponding metadata from the GitHub API.

Our paper is organized as follows: In Section \ref{section:background}, we discuss the studies that motivated our work. In Section \ref{section:rq}, we present the research questions inspired by the literature. Section \ref{section:datacuration} details the data-curation process that supports our study. Next, we describe the data-analysis process in Section~\ref{section:dataanalysis}. Section~\ref{sec:result} presents the findings for each research question. In Section~\ref{section:discussion}, we discuss our results and address threats to validity. Finally, in Section~\ref{section:conclusion}, we conclude our study.

%% file: src/background.tex
\section{Background\label{section:background}}

\mysubsection{Comprehending Dynamic Behaviors of Software}
Achieving a congruence of developers' mental models and their programs under development is a challenging but crucial task.
Gilmore studied 80 developers performing debugging tasks and found that “the success of experts at debugging is not due to better debugging skills, but rather to better comprehension”~\cite{gilmore_models_1991}. Previous studies also found that developers spent 35\%~\cite{ko_exploratory_2006} to 58\%~\cite{xia_measuring_2018} of their time on program comprehension. 
Developers often need to comprehend dynamic software behavior and align high-level functionalities with their corresponding implementation specifics. Such alignment is essential for carrying out development tasks, such as feature enhancement, debugging, and performance optimization~\cite{feng_hierarchical_2018, gilmore_models_1991, cornelissen_systematic_2009}. 

To our knowledge, we are the first large-scale study to examine ad-hoc logging practices.
The studies on how developers utilized debuggers and print statements are scattered across literature.
Some studies note similarities, such as developers’ reliance on basic tools like print statements and debuggers, rather than advanced program-comprehension tools, which are often unknown or underutilized~\cite{roehm_how_2012, cornelissen_systematic_2009}. However, there are conflicting observations regarding developers’ tool preferences. For instance, qualitative studies have reported that developers frequently use debuggers to collect runtime information~\cite{roehm_how_2012}, or even preferring them over print statements~\cite{layman_debugging_2013}. However, Beller \etal's later study on developer behaviors challenged this view, finding that ``printf debugging'', is both widely practiced and preferred by developers~\cite{beller_dichotomy_2018}.

\mysubsection{A Glimpse into Runtime Behaviors: Logs}
Logs provide developers with insights into software behavior at runtime, serving a range of purposes, from software maintenance to debugging. Unlike feature-heavy debuggers, logs are also lightweight and universal to employ~\cite{beller_dichotomy_2018}. For instance, Unix contributor Brian Kernighan maintained that ``the most effective debugging tool is still careful thought, coupled with judiciously placed print statements'' \cite{kernighan_unix_1978}. Linus Torvalds has expressed a similar view, stating that he ``never liked debuggers'' \cite{linus_lt-debugger_nodate}. However, these opinions are largely anecdotal, with little empirical evidence to substantiate them.

Beller \etal highlighted the significance of ad-hoc logs, discovering that most developers rely on print statements for debugging~\cite{beller_dichotomy_2018}. We are therefore interested in the characteristics and motivations behind these transient logs, such as their locations, content, and developers’ underlying intentions. However, to our knowledge, only one study by Jiang \etal~\cite{jiang_log-it_2023} has explicitly focused on ad-hoc logging, while other studies have primarily analyzed production logs for system maintenance. We categorize these studies into four areas, which we label as: \textbf{Why}, \textbf{How}, \textbf{Where}, and \textbf{What}.

\textbf{Why}: Li \etal surveyed 66 developers and analyzed 223 logging-related issue reports, identifying four key benefits of logging: troubleshooting support, execution tracking, aiding comprehension, and bookkeeping~\cite{li_qualitative_2021}. 
In the context of embedded software engineering, Yang \etal interviewed 28 developers and found that production logs are mainly used for problem localization and performance improvement, rather than understanding or reverse engineering existing systems~\cite{yang_interview_2021}.

\textbf{How}: Jiang \etal interviewed seven developers as formative research for their ad-hoc log visualization tool, identifying four key challenges in log authoring~\cite{jiang_log-it_2023}: lack of meaningful organization in log output, insufficiently informative data visualizations, frequent context loss due to view switching, and trade-offs in crafting and interpreting logs. Li \etal, focused on logging \emph{levels} and found that existing logging statements and containing block types are the major factors in determining the appropriate log level for the newly added log statements~\cite{li_which_2017}. 

\textbf{Where}: The location of production logs has been widely studied. Fu \etal conducted a mixed-methods study at Microsoft, combining code analysis on two large systems with interviews, and identified five categories of log locations and their distributions~\cite{fu_where_2014}. Similarly, Li \etal analyzed seven large open-source Java systems, identifying seven categories of log locations and corresponding distributions~\cite{li_where_2020}. Both studies developed log-location recommendation systems, laying the groundwork for future tools~\cite{mastropaolo_using_2022, zhu_learning_2015, li_towards_2020}.

\textbf{What}: The content of log statements has received less focus. Liu \etal developed a recommendation tool using RNNs to suggest which variables developers should log, though the specific characteristics of these variables were not detailed~\cite{liu_which_2021}. Jiang \etal observed that developers often use template strings and labels to facilitate context switching between source code and log output~\cite{jiang_log-it_2023}.

These studies, mostly focusing on production logs, have informed our research questions and provided methodological insights and points of comparison for our work.

\mysubsection{Capturing Developers Glimpses through Mistakes}
We hypothesize that limited research on comprehending software’s dynamic aspects stems from the challenges of recruiting participants and analyzing their data. To capture often-hidden traces like ad-hoc logs from developers' local environments, studies have traditionally relied on lab setups~\cite{ko_practical_2015}, video analysis of small groups~\cite{alaboudi_exploratory_2021}, or large-scale surveys to approximate developer behaviors~\cite{maalej_comprehension_2014}. For example, Maalej \etal surveyed 1,447 participants to identify comprehension-supporting factors and information sources, such as documentation~\cite{maalej_comprehension_2014}. Alaboudi \etal analyzed 15 live-stream videos to study the patterns of debugging, editing, and execution cycles~\cite{alaboudi_exploratory_2021, alaboudi_edit_2021}. Similarly, Beller \etal recruited 458 developers to install custom IDE plugins to monitor debugging practices~\cite{beller_dichotomy_2018}. Despite these contributions, replicating or building on such studies remains difficult due to the complexities and costs of human-subject research—challenges that include recruitment difficulties~\cite{beller_how_2015}, contextual variability~\cite{lung_difficulty_2008}, ethical considerations, and potential biases~\cite{ko_designing_2004}. For instance, Beller \etal reported that after 60 retweets of their study invitation, only two new participants joined~\cite{beller_how_2015}. Additionally, even with recruited developers, it remains difficult to differentiate ad-hoc logs from those intended for production when monitoring their typing in IDE.

The challenges of working with human subjects for incremental studies led us to draw on previous mining software repositories (MSR) research, especially for those works that uncover hidden patterns in revision histories. For example, Beyer \etal identified bug-fixing patterns from GitHub histories~\cite{beyer_p3_2024}, and Sinha \etal published data on secret key leaks caused by developers' mismanagement of public version control systems~\cite{sinha_detecting_2015}. Observing developers’ frequent mistakes in version control, we mined and published ad-hoc logs by examining patches made after inadvertent changes. Although it is possible that developers could \emph{intentionally} commit such logs, we consider them as mistakes for multiple reasons: (1) we cannot know the initial developer intent and the subsequent deletion may indicate a fix, (2) \textit{console.log} instructions in JavaScript output to transient consoles for users, and (3) community guidelines (\eg Airbnb~\cite{noauthor_airbnbjavascript_2025}) and widely used linters (\eg ESLint~\cite{linter}) discourage committing \textit{console.log} statements to version-control systems. 
% \yihung{While some of the deleted logs might be pushed by developers intentionally for future debugging purposes, we treated these logs as mistakes because community guidelines (e.g., Airbnb) and widely used linters (e.g., ESLint) discourage committing console.log statements to version-control systems.}

To gain deeper insights into the motivations behind these logs, we also conducted qualitative analysis on Alaboudi \etal's video data ~\cite{alaboudi_exploratory_2021}. We hope our exploratory study and dataset can inspire further research on ad-hoc logging practices and stimulate more in-depth discussions on developers' comprehension of software’s dynamic behaviors.

%% file: src/dataset.tex
\section{Research Questions~\label{section:rq}}
Built upon previous research in understanding \emph{Where}, \emph{What}, \emph{Why}, and \emph{How}, we formulate the below research questions for analyzing our collected data.

\begin{description}

    \item[RQ1:] \emph{Where} in code do developers place ad-hoc logs? 
    \item[RQ2:] \emph{What} information do developers include in ad-hoc logs? 
    \item[RQ3:] \emph{Why} do developers use ad-hoc logs?
    \item[RQ4:] \emph{How} do developers manipulate (insert, revise, and remove) ad-hoc logs to achieve their goals? 

\end{description}
\begin{table*}[t]
    \caption{Explanation of Fields and Examples for the Logs Metadata\label{tab:crawled_metadata}}
    \centering
    \begin{tabularx}{\textwidth}{@{}p{0.25\textwidth}X@{}}
        \toprule
        \textbf{Field} & \textbf{Description and Example} \\ \midrule
        
        \texttt{logInString} & \parbox[t]{\textwidth}{The string content of the log statement.\\ Example: \texttt{"console.log(pattern)"}} \\ \midrule
        
        \texttt{functionName} & \parbox[t]{\textwidth}{The name of the function where the log resides.\\ Example: \texttt{"calcAreasByPattern"}} \\ \midrule
        
        \texttt{functionType} & \parbox[t]{\textwidth}{The type of the function (e.g., FunctionDeclaration).\\ Example: \texttt{"FunctionDeclaration"}} \\ \midrule
        
        \texttt{logLoc} & \parbox[t]{\textwidth}{The location of the log in the code, including the start and end positions (line and column).\\ Example: \texttt{\{"start": \{"line": 95, "column": 4\}, "end": \{"line": 95, "column": 24\}\}}} \\ \midrule
        
        \texttt{complexityOfFunction} & \parbox[t]{\textwidth}{Metadata about the complexity of the function.\\ Example: \texttt{\{"name": "calcAreasByPattern", "complexity": 3, "line": 87\}}} \\ \midrule
        
        \texttt{arguments} & \parbox[t]{\textwidth}{A list of the arguments passed to the function.\\ Example: \texttt{[ \{"str": "pattern", "typeOfArg": "Identifier"\} ]}} \\ \midrule
        
        \texttt{isAsyncFunction} & \parbox[t]{\textwidth}{A boolean indicating whether the function is asynchronous.\\ Example: \texttt{false}} \\ \midrule
        
        \texttt{isCallbackFunction} & \parbox[t]{\textwidth}{A boolean indicating whether the function is a callback function.\\ Example: \texttt{false}} \\ \midrule
        
        % \texttt{linesOfCodeBetweenLogAndWrapperFunc} & \parbox[t]{\textwidth}{The number of lines of code between the log statement and the enclosing function.\\ Example: \texttt{8}} \\ \midrule
        
        \texttt{isAnonymousFunction} & \parbox[t]{\textwidth}{A boolean indicating whether the function is anonymous.\\ Example: \texttt{false}} \\ \midrule
        
        \texttt{blockStatement} & \parbox[t]{\textwidth}{The type of block statement where the log resides, such as FunctionDeclaration, TryStatement.\\ Example: \texttt{"FunctionDeclaration"}} \\ \midrule
        
        \texttt{repositoryName} & \parbox[t]{\textwidth}{The name of the repository where the code is stored.\\ Example: \texttt{"0067ED\_vue-block"}} \\ \midrule
        
        \texttt{commitSha} & \parbox[t]{\textwidth}{The commit SHA of the repository at the time of removing the logs.\\ Example: \texttt{"f0ceff46bc35c9caad200fcbc4d53892c5a966a6"}} \\ \midrule
        
        \texttt{folderPath} & \parbox[t]{\textwidth}{The file path to the folder where the log is located within the repository.\\ Example: \texttt{"src\_components\_algorithm\_area\_js"}} \\

        \bottomrule
    \end{tabularx}
\end{table*}

\begin{table*}[t]
    \vspace{-.1in}
    \caption{Explanation of Fields and Examples for the Repositories Metadata~\label{tab:repo_info}}
    \centering
    \begin{tabularx}{\textwidth}{@{}p{0.25\textwidth}X@{}}
        \toprule
        \textbf{Field} & \textbf{Description and Example} \\ \midrule

        \texttt{description} & \parbox[t]{\textwidth}{The description of the repository\\ Example: \texttt{Independent technology for ...}} \\\midrule
 
        \texttt{full\_name} & \parbox[t]{\textwidth}{The name of the owner and the name of the repository\\ Example: \texttt{TryGhost/Ghost}} \\\midrule
    
        \texttt{contributors, stars, forks\_count, watchers\_count, size} & \parbox[t]{\textwidth}{Basic statistics of the number of contributors, stars, forks, watchers, and size, of the repositories.\\ Example: \texttt{1154693}} \\ \midrule

        \texttt{lastUpdated} & \parbox[t]{\textwidth}{Last updated time when we query the repository on October 29, 2024\\ Example: \texttt{2016-03-04T08:20:37Z}} \\\midrule
        
        \texttt{isActive} & \parbox[t]{\textwidth}{Whether the repository is actively maintained in the last 6 months when querying\\ Example: \texttt{false}} \\

        \bottomrule
    \end{tabularx}
    \vspace{-.1in}
\end{table*}

\section{Dataset Curation\label{section:datacuration}}
% \begin{figure}
%     \centering
%     \includegraphics[width=1\linewidth]{src/figures/data extraction process.pdf}
%     \caption{Enter Caption}
%     \label{fig:enter-label}
% \end{figure}
\begin{figure}[t]
    \centering
    \includegraphics[width=\linewidth]{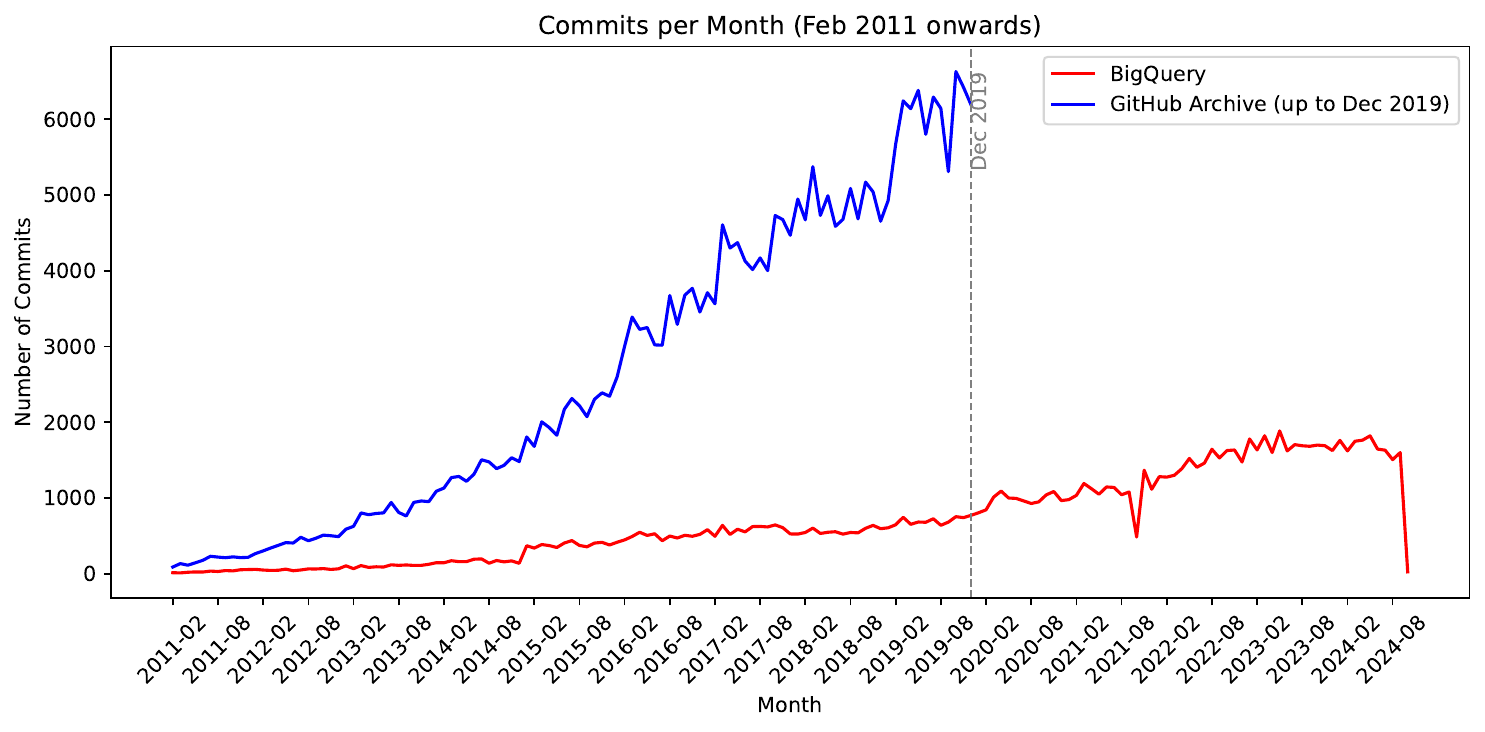}
    \caption{Two Data Sources (GitHub Archive \& Google BigQuery) and Their Distributions}
    \label{fig:datasources}
\vspace{-5pt}
\end{figure}
% provide an overview to discuss how do we crawl the data from github and also how do we curate the video data
In this section, we provide an overview of the curation process for two types of data that enabled our empirical study: (1) ad-hoc logs and their corresponding metadata, such as commit, repository metadata, before-and-after code; and (2) live-stream videos of developers inserting ad-hoc logs for debugging or comprehension. To ensure reproducibility, the complete dataset and crawling scripts are available in \cite{logs-commits}.

\mysubsection{Ad-hoc Log Statements\label{subsec:AdhocLogDataCuration}} 
To collect logs that were later removed by developers in JavaScript, we first identified the targeted commits from GitHub Archive. The process began by extracting all events of type \emph{PushEvent} from GitHub Archive. From these events, we extracted the commits whose messages matched the target regular expression:  \emph{(remove$\vert$ delete).%
*?%
console.log}.

Due to the rate limitations of the GitHub API, we gathered \emph{PushEvent}s from February 12, 2011, to January 11, 2020. These events could contain commits that dated back to 2007. Later, we supplemented this dataset with a sample of additional commits through Google BigQuery, covering the period up to October 1, 2024. The SQL query used in BigQuery selected only the first commit from each \emph{PushEvent} from the entire GitHub Archive and filtered them using the same regular expressions. Figure~\ref{fig:datasources} illustrates the distribution of data from these two sources.

After retrieving the commits, we used GitHub APIs to collect the necessary data that is missing from the event metadata, such as number of stars of the repository, for subsequent analysis. 
The structure of the data that we gathered for each log and each repository are shown in are provided in Table~\ref{tab:crawled_metadata} and Table~\ref{tab:repo_info}, respectively. 
The crawling scripts and dataset are available in~\cite{logs-commits}. We collected a total of 364,837 unique commits, with 276,927 sourced from the GitHub Archive and 87,910 from the GitHub BigQuery API sample. Of these, 30,529 commits were found in both sources. The average number of commits per repository is 1.54 and the average number of commits per month is 1,745.63.
%\jim{I was expecting to see some information about the projects that you mined. At least, at first, I was expecting to see project names (as that is what I typically see in conferences like ICSE). But, I now understand that the number of projects is probably WAY too big to list. So, then, I'm surprised that I don't see some impressive statistic about how many projects were mined. Just like, a number of projects that informed your data.}
We collected a total of 202,144 repositories from GitHub Archive, of which we can use 156,265 repositories because they are available through GitHub API.
Such discrepancy may be due to repositories that were either deleted or made private.

Figure~\ref{fig:cumulativeProportion} shows the cumulative ratio of projects that had any activity during the last N months.
We found most projects (81.4\%) showed no recorded activity within the six months prior to our data collection on October 29, 2024. 
Table~\ref{tab:stats_of_repos} shows the basic statistics of the number of contributors, stars, forks, watchers, and size. We could see that most of them are lean toward small number, indicating most of the collected repositories are personal and small.
% \begin{table}[t]
%     \caption{Statistical summary of repository metrics.\label{tab:stats_of_repos}}
%     \label{tab:stats}
%     \centering
%     \begin{tabular}{lccccc}
%         \toprule
%         Metric & Mean & Std Dev & Min & Median & Max \\
%         \midrule
%         Stars                     & 206.71 & 2978.04 & 0 & 0 & 403,857 \\
%         Forks                     & 36.80 & 524.81 & 0 & 0 & 71,736 \\
%         Watchers                  & 206.71 & 2978.04 & 0 & 0 & 403,857 \\
%         Size (bytes)              & 63,012.31 & 704,317.70 & 0 & 2,417 & 55,403,320 \\
%         Contributors              & 43.81 & 299.58 & 1 & 3 & 21,163 \\
%         \bottomrule
%     \end{tabular}

% \end{table}
% \begin{figure}[tb]
%     \centering
%     \includegraphics[width=0.99\linewidth]{src/figures/cumulative_proportion_plot.pdf}
%     \caption{Cumulative Proportion of Projects by Months Since Last Updated for All Collected Repositories \april{can we make the labels larger to read?} }
%     \label{fig:enter-label}
% \end{figure}

\mysubsection{Live-Streamed Videos}
In this subsection, we detail our data-curation process, building on the work of Alaboudi and LaToza~\cite{alaboudi_what_2023}, where researchers observed developers live streaming their software development and documented their working cycles.
Developers, in particular, are adopting live streaming to increase visibility, facilitate education, and foster community~\cite{kokinda_streaming_2023}.
Such a dataset provides unobtrusive observation of behaviors in natural settings, which can reduce observer bias~\cite{mahtani_catalogue_2018}. Despite its potential, live streaming remains underexplored in software-engineering research. Fortunately, prior work by Alaboudi and LaToza\cite{alaboudi_what_2023} has made significant strides by collecting and labeling live-streamed videos and developing the \emph{observer-dev.online} platform~\cite{online-dev}, tailored for software-engineering researchers. We used all 15 videos (total duration: 36:01:30) that they curated and conducted a thematic analysis, with both the dataset and our qualitative findings made publicly available \cite{qualitative-data-analysis}. While little research has explored why and how developers insert log statements, Alaboudi and LaToza~\cite{alaboudi_what_2023} segmented each video into debugging episodes, such as \textit{“Testing the program and reading the outputs”} or \textit{“Interacting with a file of code (Edit).”} These segments allowed us to efficiently pinpoint moments when developers began adding log statements to their source code.

%% file: src/empirical_study.tex
\section{Data Analysis~\label{section:dataanalysis}} 
In this section, we detail our data-analysis process that enable answering the above research questions.
\begin{figure}[t]
    \centering
    \includegraphics[width=\linewidth]{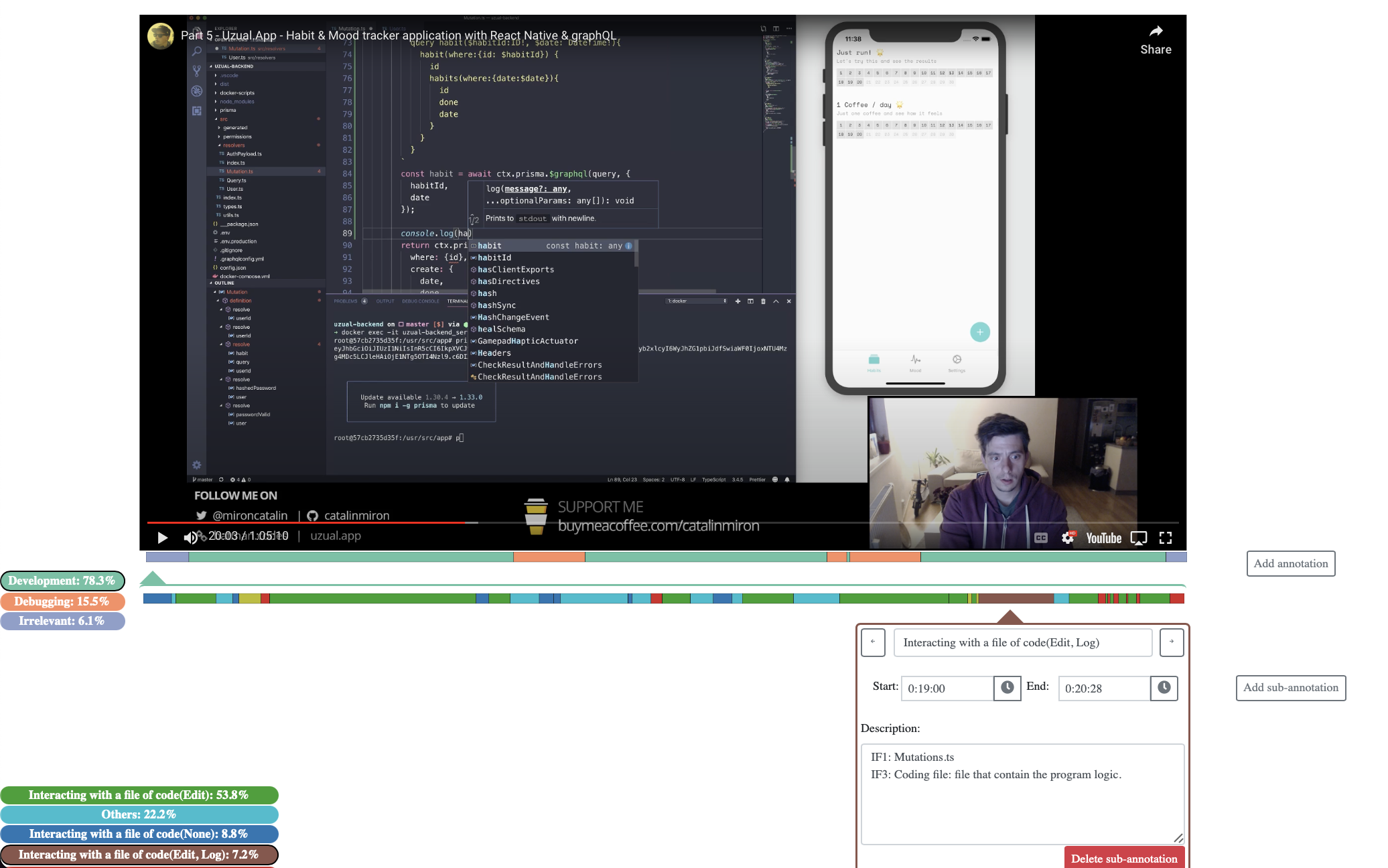}
    \caption{Screenshot of Observe-dev.online, when a developer is adding a log into their code}
    \label{fig:screenshot_of_observer_online}
\end{figure}
\subsection{Analyzing Ad-hoc Logs through Mistakes\label{sec:ad-hoc-logs-analysis}}

To address RQ1 and RQ2, we prepared the dataset for analysis by conducting the following preprocessing steps:

\textbf{Filtering}: Developers sometimes include commit messages without corresponding file changes or add files outside our target programming languages. To address this, we generated diff files by comparing only JavaScript and TypeScript files before and after each commit. Additionally, we filtered out library files and minified files, using Google Chrome DevTools' minified file-detection algorithm~\cite{google-chrome}. We then processed the \emph{before} files to search for \emph{console.log} \emph{CallExpressions}. When located, we extracted these logs if the corresponding line number matched the deleted lines in the diff files.

\textbf{Contextualizing Log Messages}: Extracting log messages alone lacks context. We predefined metadata for static analysis, capturing the functions and blocks containing each log, the parameter types in log statements, and Cyclomatic Complexity~\cite{ebert_cyclomatic_2016} using the \emph{cyclomatic-complexity} library~\cite{cyclomatic}. The scripts and metadata are accessible at \cite{logs-commits}. To identify removed lines of code, we analyzed diff files for deleted \emph{console.log} lines. Using the \emph{@typescript-eslint/typescript-estree} library, we parsed the AST, traversing from the \emph{CallExpression} node to identify enclosing blocks and functions, and inferred names for anonymous functions based on Figure~\ref{fig:customized_function_names}. From 364,837 commits, 215,160 contained removed logs. A single-threaded analysis, run over two days, extracted 548,880 ad-hoc logs, stored in a 436.7 MB JSON file.

\textbf{Preparing samples for comparison}: To estimate complexity and function-type distribution, we prepared three repository samples. The first sample includes 1,000 JavaScript repositories created between 2023 and 2024, randomly selected from approximately 3,000 results using the GitHub Search API. The second and third samples each contain 1,000 repositories, randomly selected from GitHub Archive and GitHub BigQuery. Detailed distributions are provided in Section~\ref{sec:result}.
    
\begin{table}[t]
    \caption{Statistical summary of repository metrics.\label{tab:stats_of_repos}}
    \label{tab:stats}
    \centering
    \begin{tabular}{lccccc}
        \toprule
        Metric & Mean & Std Dev & Min & Median & Max \\
        \midrule
        Stars                     & 206.71 & 2978.04 & 0 & 0 & 403,857 \\
        Forks                     & 36.80 & 524.81 & 0 & 0 & 71,736 \\
        Watchers                  & 206.71 & 2978.04 & 0 & 0 & 403,857 \\
        Size (bytes)              & 63,012.31 & 704,317.70 & 0 & 2,417 & 55,403,320 \\
        Contributors              & 43.81 & 299.58 & 1 & 3 & 21,163 \\
        \bottomrule
    \end{tabular}
    \vspace{-.1in}
\end{table}
\begin{figure}[tb]
    \centering
    \includegraphics[width=0.99\linewidth]{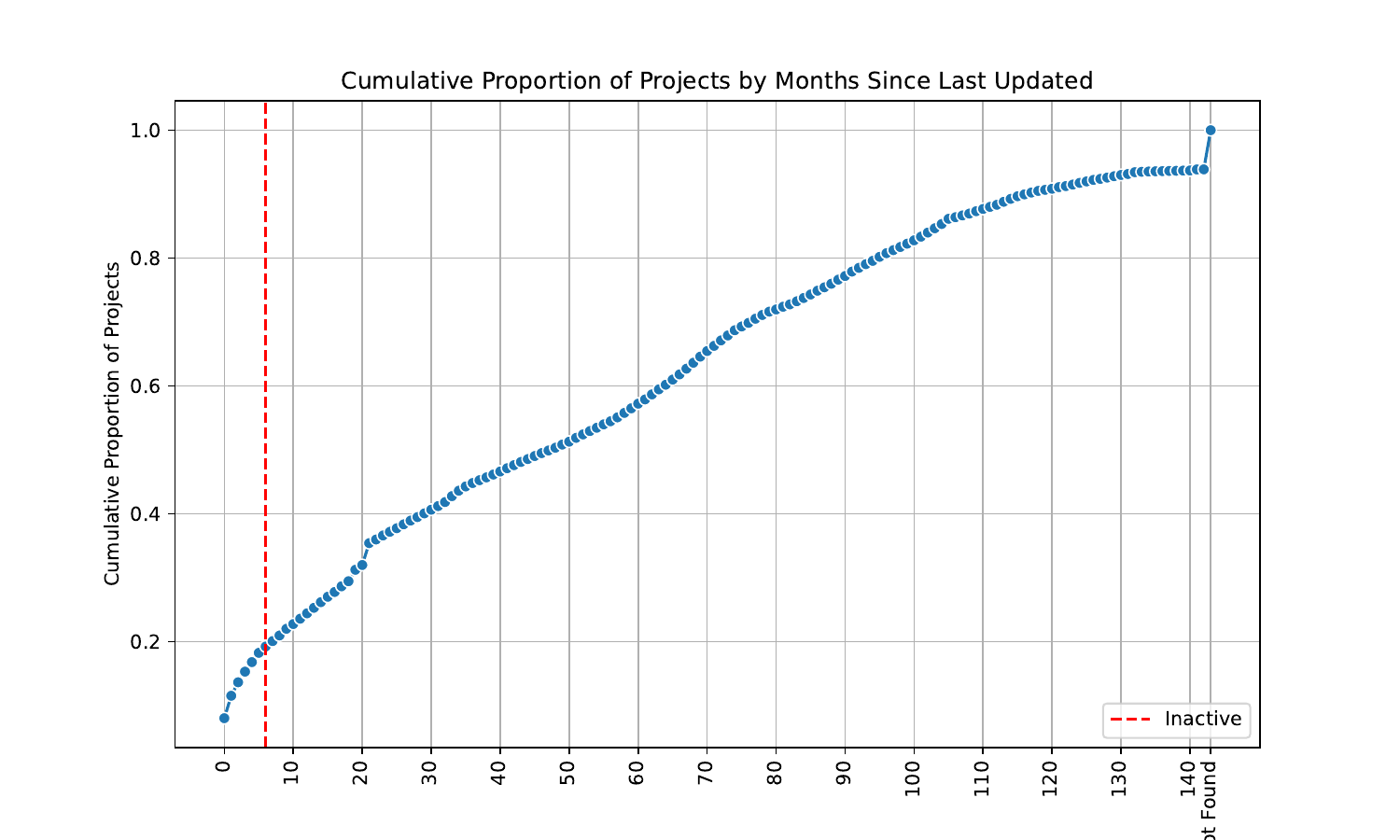}
    \vspace{-.1in}
    \caption{Cumulative Proportion of Projects by Months Since Last Updated for All Collected Repositories.}
    \label{fig:cumulativeProportion}
    \vspace{-.2in}
\end{figure}

A concern with our approach is whether the removed logs were genuinely ad-hoc. To validate this, we manually analyzed 100 sampled commits. Logs were labeled as ad-hoc if they were not replaced by other logs or by functions from third-party logging libraries. Among 130 files with 269 removed logs, only 5 logs (1.86\%) were labeled as non-ad-hoc. The labeled data is available in~\cite{manual-commits-label}.

\begin{figure}[t]
    \footnotesize
    \begin{align*}
\mathrm{Variable~Declarator} &: \mathtt{var~\textcolor{red}{cb}~=~function~()~=>~\{ \ldots \}} \\
\mathrm{Call~Expression}     &: \mathtt{fetch(\ldots).\textcolor{red}{then}(()~=>~\{ \ldots \})} \\
\mathrm{Assignment~Expression} &: \mathtt{\textcolor{red}{cb}~=~function()~=>~\{\ldots\}} \\
\mathrm{Class~Property} &: \mathtt{class~Text~\{ \textcolor{red}{getText}~=~()~=>~\{\ldots\}~\}}\textbf{}
    \end{align*}
    \caption{Custom approach to assign names to anonymous functions. Tokens highlighted in red represent the names assigned to these functions. We use such method to conduct analysis on function names later.} 
    \label{fig:customized_function_names}
    \vspace{-.1in}
\end{figure}

% \begin{figure}[tb]
%     \begin{verbatim}
% Variable Declarator: var cb = function () => { ... }
%     Call Expression: fetch(...).then(() => { ... }
%     \end{verbatim}
% \end{figure}
\begin{table*}[t]
    \caption{Distributions of Different Type of Functions from Three Samples~\label{tab:distribution_of_function_type}}
    \label{tab:my_label}
    \centering
    \begin{tabular}{llllll}
    \toprule
         Sample Type&  Total Function No. & Async Function (\%)&  Anonymous Function (\%)&  Functions as Direct Callback& Others\\
    \midrule
         Sample 2023-2024 & 1,070,838 &  46,940 (4\%)& 664,649 (62\%)&  249,743 (23\%)& 377,939 (35\%) \\ 
         Sample GitHub Archive & 2,937,684 & 18,628 (0.63\%) & 2,393,852 (81.49\%) & 902,542 (30.72\%) & 516,581 (17.58\%)\\
         Sample GitHub BigQuery & 2,009,471 &  146,351 (7.28\%) & 1,725,954 (85.89\%) & 771,356 (38.39\%) & 262,625 (13.06\%)\\
    \bottomrule
    \end{tabular}
    \vspace{-.1in}
\end{table*}

\subsection{Qualitative Analysis of Live Streaming Data}
To contextualize the ad-hoc logs for answering RQ3 and RQ4, we began the analysis of live-streaming data by identifying the debugging episodes within the labeled data created by Alaboudi \etal~\cite{alaboudi_what_2023}, specifically those marked as “interacting with a file of code (Edit, Log)” or “interacting with a file of code (Log).” For each identified video segment, we reviewed it by either rewinding or fast-forwarding until we can address a series of pre-crafted guiding questions for our deductive thematic analysis~\cite{braun_thematic_2012}.
Fig~\ref{fig:screenshot_of_observer_online} shows a screenshot of observe-dev.online, when a developer is inserting a log into their codebase. We answered the following questions by cross-referencing streamers' behaviors and their verbal expressions.
\begin{itemize} 
    \item How and why did they insert the log statement? 
    \item When and how did they remove the log statements? 
    \item Where and what exactly did they log? 
\end{itemize} 

The first author initially coded five videos containing 45 ad-hoc logs and developed a codebook, which is available in the replication package~\cite{qualitative-data-analysis}. The second author then independently coded the same five videos. There was substantial agreement between the two researchers, as indicated by Cohen’s $\kappa = 0.849$. Any disagreements that arose were discussed with the research team during weekly meetings until a consensus was reached. Subsequently, one researcher proceeded with coding the remaining videos. 
A total of fifteen videos from~\cite{alaboudi_what_2023}, comprising over 36 hours of content, were analyzed to identify 92 ad-hoc logs. Each video was labeled from V1 to V15. Three videos contained no ad-hoc logs. For instance, V5 and V6, programmed in Rust, relied solely on compile-time error logs for debugging, while V14, programmed in C\#, consistently used the Visual Studio debugger.

%%%%%%%%%%%%%%%%%%%%%%%%%%%%%%%%%%%%%%%%%%%%%%%%%%%%%%%%%%%%%%%%%%%%%%

\section{Results~\label{sec:result}}

Using the methodology described in the prior two sections, we produced a number of results and findings, which we present here, organized by the research questions that we seek to answer.

\subsection{\textbf{RQ1:} Where do developers put these logs in their code?}

\begin{figure}[tb]
    \centering
    \includegraphics[width=1\linewidth]{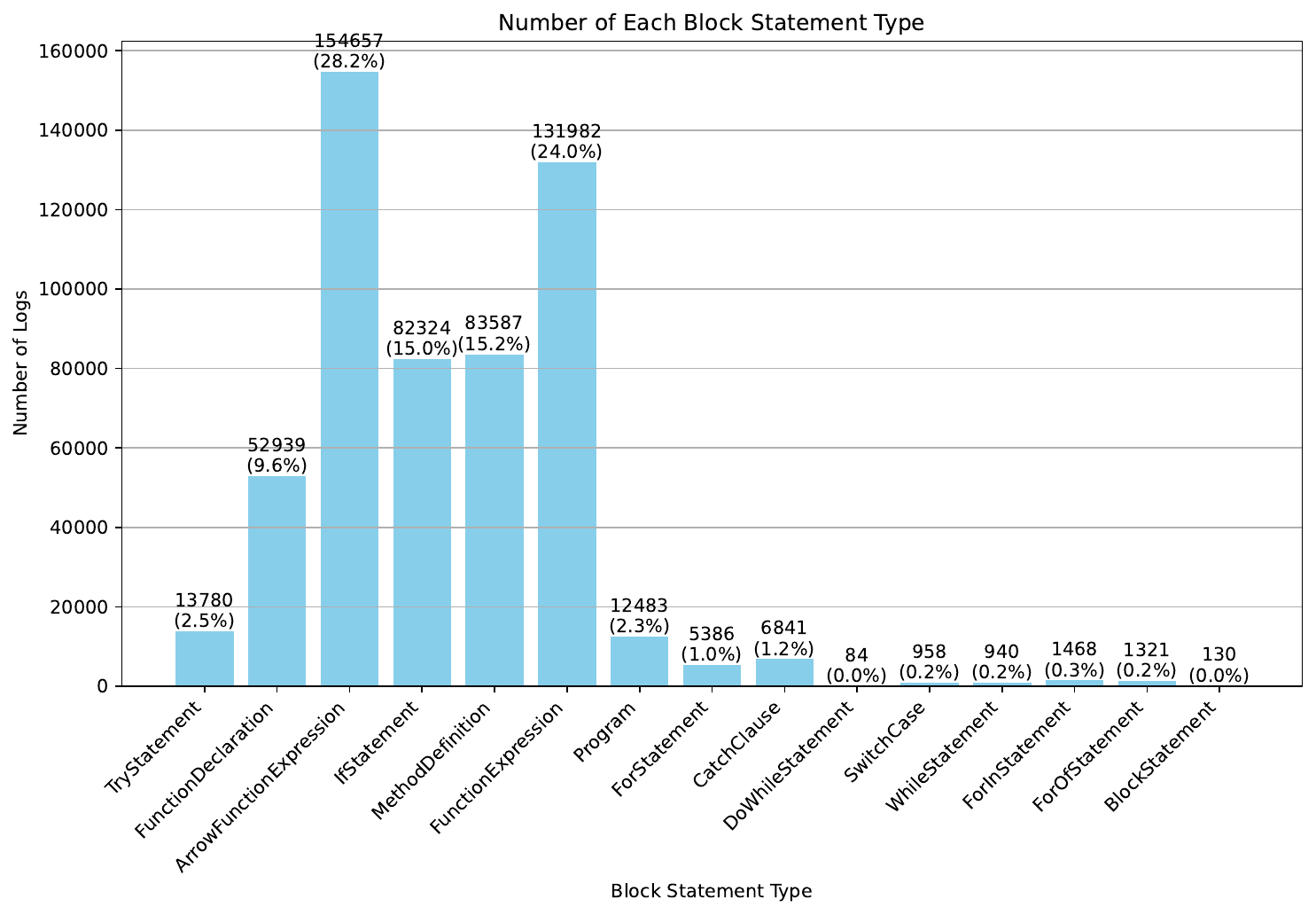}
    \vspace{-.2in}
    \caption{Types of Block where the Ad-hoc Resides}
    \label{fig:distribution_of_block_statements}
    \vspace{-.2in}
\end{figure}

\textbf{Distribution of Ad-hoc Logs vs. Logs in Production.} 
Figure \ref{fig:distribution_of_block_statements} shows the results of where developers mistakenly placed ad-hoc logs, based on our analysis from both sources. The largest proportion of these ad-hoc logs appears within ArrowFunctionExpressions, which use a shorter function syntax and account for 28.2\% of these placements. Additionally, 24.0\% of logs are placed within FunctionExpressions, 9.6\% are found in FunctionDeclarations, and 15.2\% are found in MethodDefinition.

Both FunctionExpression and FunctionDeclaration represent functions created with the function keyword. However, FunctionExpressions create functions in expressions, which may or may not have names, whereas FunctionDeclarations create named functions directly. The MethodDefinition is a shorter syntax for defining a function property in an object. All represent root-level placements within a function (\ie not in a deeper block, such as an if block) in any form, which accounts for 77\% in total. Also notice that TryStatement and CatchClause placements only account for 2.5\% and 1.2\% of the placements, respectively. 

These results differ from prior works' reporting of the location of \emph{production logs}, which showed that 69.9\% of the logs were either in error catch blocks or right after if statements~\cite{li_where_2020}.

Our qualitative analysis of live-streamed coding videos reveals a similar distribution: 71 logs (77.17\%) out of 92 appear in function bodies, with only one located in a catch block.
We believe this alignment reaffirms the validity of our dataset and highlights a fundamental difference in log motivations: logs in production aim to support maintenance, while ad-hoc logs primarily aid program comprehension.

\vspace{5pt} 
\noindent\fbox{\begin{minipage}{0.47\textwidth}
\textbf{Finding 1A}:
Most ad-hoc logs were placed at the root level of function blocks (77\%), which differs from the distribution found in production logs, where half are located in catch or branch blocks~\cite{li_where_2020, fu_where_2014, yuan_characterizing_2012}.
\end{minipage}}
\vspace{5pt} 

\textbf{Types of Functions with Ad-hoc Logs.}
Understanding a JavaScript system can be challenging for developers due to the wide use of asynchronous and event-driven functions. This challenge has led to extensive research in visualization and static analysis to help developers debug these specific features more effectively~\cite{alimadadi_understanding_2016, turcotte_drasync_2022}. 
As such, we are curious to study if such functions (specifically anonymous, callback, and asynchronous) received more use of ad-hoc logs for developers to understand them at runtime. 
To assess the overall distribution of these function types as a baseline,
we followed the sampling method described in Section~\ref{sec:ad-hoc-logs-analysis}, and present the distributions of different types of the functions from three samples in Table~\ref{tab:distribution_of_function_type}. We can see the majority of the functions are anonymous, asynchronous function is only account for less than 10\%, and callbacks are less than 40\%.

\begin{figure}[tb]
    % \vspace{-.1in}
    \centering \includegraphics[width=1\linewidth]{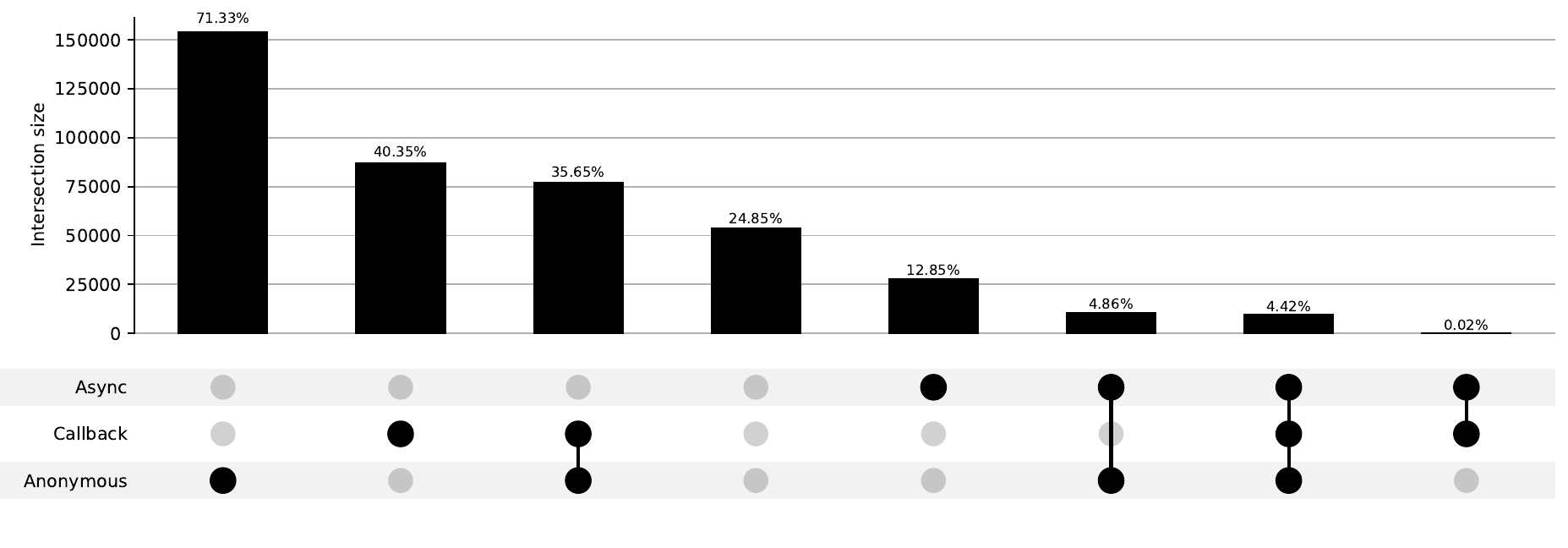}
    \caption{UpSet~\cite{lex_upset_2014} Diagram with Intersect Mode of Different Types of the Functions Where Ad-hoc Logs Resides}
    \label{fig:UpSet}
    \vspace{-.1in}
\end{figure}
\begin{figure}
    \centering
    \vspace{-.1in}
    \includegraphics[width=1\linewidth]{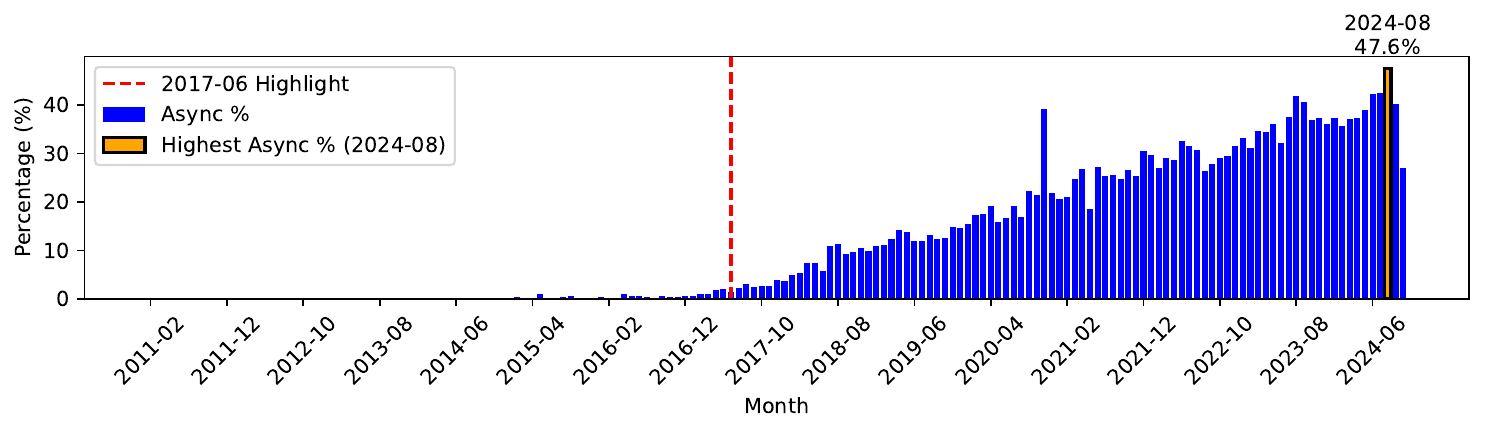}
    \vspace{-.1in}
    \caption{Percentage of Ad-hoc Logs within Asynchronous Functions over Time.~\label{fig:async_func}}
    \label{fig:enter-label}
\end{figure}
We present the distribution of function types where ad-hoc logs were placed in Figure~\ref{fig:UpSet}, which shows the distributions of function types across \textbf{Async}, \textbf{Callback}, and \textbf{Anonymous} functions. \textbf{Async} functions are marked with the \emph{async} keyword and are commonly used to handle asynchronous operations, such as HTTP requests.
\textbf{Callback} functions, in this context, refer to functions defined directly within the call site of another function, as it is impractical to exhaustively determine if all functions are passed as parameters.
\textbf{Anonymous} functions are defined without names.

Additionally, we analyzed how the distribution of function types has changed over time. We present only the changes in the distribution of \textbf{Async} functions in Figure~\ref{fig:async_func}, as the distributions of \textbf{Callback} and \textbf{Anonymous} functions remain relatively stable over time; details for these can be found in the replication package~\cite{logs-commits}. Figure~\ref{fig:async_func} shows that the percentage of asynchronous functions containing ad-hoc logs has gradually increased since the introduction of the \emph{async} keyword around mid-2017, reaching as high as 47.6\% in August 2024. This percentage is significantly higher than in our 2023--2024 sample repositories, where only 4\% of collected functions are asynchronous. Similarly, the qualitative analysis shows that in all JavaScript videos, 45.3\% (24 out of 53) of the ad-hoc logs are in the \textbf{Async} and \textbf{Callback}.

\begin{figure}[b]
    \vspace{-.1in}
    \centering
    \includegraphics[width=\linewidth]{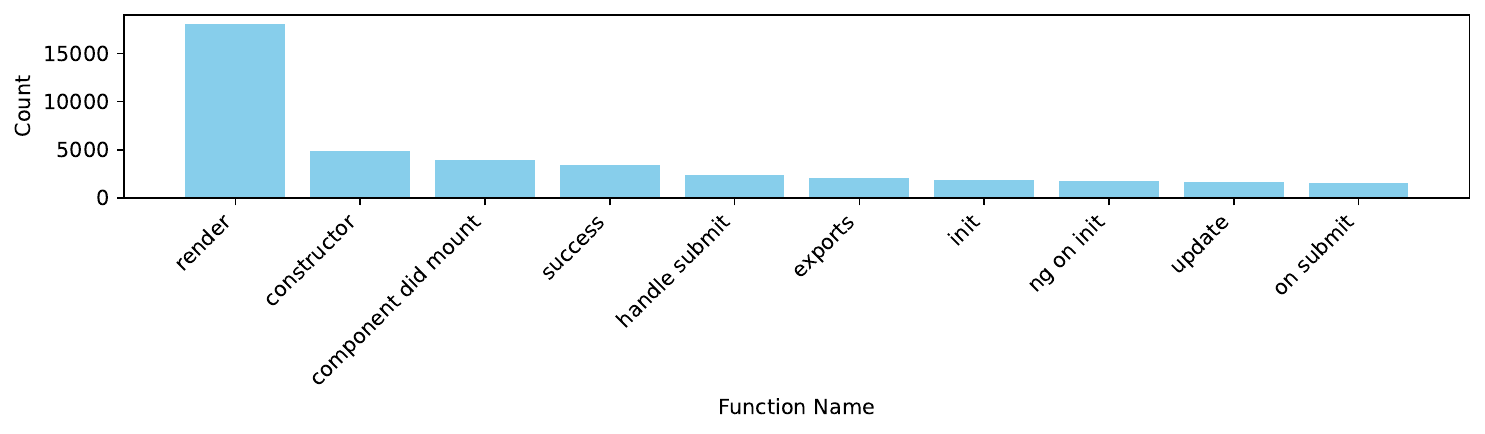}
    \vspace{-.2in}
    \caption{Top 10 Most Frequent Function Names Where Ad-hoc Logs Reside}
    \label{fig:top_10_most_frequent_function}
    \vspace{-.1in}
\end{figure}
\begin{figure}[b]
    \centering
    \includegraphics[width=\linewidth]{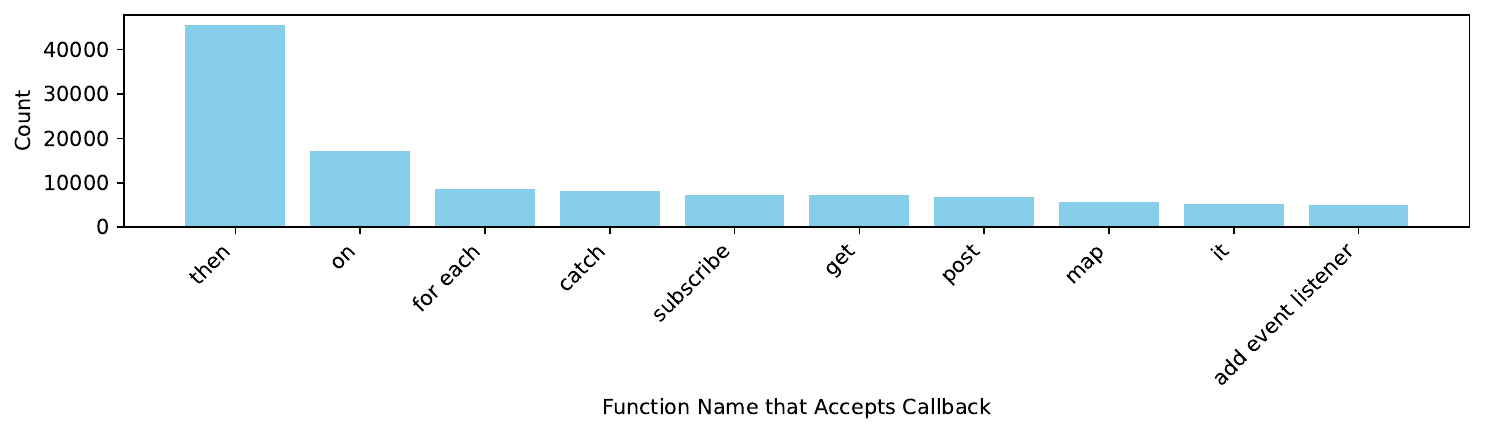}
    \vspace{-.2in}
    \caption{Top 10 Most Frequent Function Name that Accepts Callbacks Where Ad-hoc Logs Reside}
    \label{fig:callback_name}
    \vspace{-.1in}
\end{figure}

\textbf{Name of the functions with ad-hoc logs.}
We are also interested in two specific types of functions: the names of functions that accept other functions containing ad-hoc logs as arguments, and the names of functions or methods extracted using our custom naming approach, as defined in Figure~\ref{fig:customized_function_names}.
Figure 9 presents the top-10 most frequent function and method names in which ad-hoc logs were found. As shown, \emph{render} is the most common name, appearing over three times more frequently than the second-most common name. This result suggests that ad-hoc logs are often placed in lifecycle functions commonly seen in frontend libraries like React or Vue. Similarly, \emph{componentDidMount}, \emph{init}, and \emph{ngOnInit} are lifecycle functions frequently used in React, Angular, and Vue. Our qualitative data also shows that the developer in V13 placed an ad-hoc log to understand the React lifecycle, where the timing is controlled internally by React, as they noted, “\textit{My question is why does it render once at all.}”

Figure~\ref{fig:callback_name} displays the top-10 most frequent names of functions that accept callbacks. The most common name, \emph{then}, appears over twice as often as the second-most common name. This pattern suggests that ad-hoc logs were frequently placed within callback functions accepted by the \emph{then} function. The distribution implies that many callbacks support asynchronous behaviors, as \emph{then} is typically associated with Promises, commonly used for handling asynchronous operations before the introduction of the \emph{async} keyword.

\vspace{5pt} 
\noindent\fbox{\begin{minipage}{0.47\textwidth}
\textbf{Finding 1B}: Many ad-hoc logs are placed into callbacks, asynchronous, and library lifecycle functions. Moreover, the proportion of these logs in asynchronous functions increased from its introduction in 2007 to 2024. These findings suggest that developers use ad-hoc logs to help them in their struggles to understand how such indirect control-flow actually work at runtime.
\end{minipage}}
\vspace{5pt} 

\textbf{Complexity of the function with ad-hoc logs.}
We suspected that function complexity may influence the placement of ad-hoc logs. To investigate, we compared the cyclomatic complexity of functions containing ad-hoc logs with the complexity of all functions in the sample repositories from both BigQuery and GitHub Archive sources. Our analysis revealed significant differences in mean complexity in both datasets. For functions with ad-hoc logs from BigQuery ($\mu = 7.77$, $\sigma = 19.19$, $\tilde{x} = 4.0$), the mean complexity was higher than that of all functions ($\mu = 3.32$, $\sigma = 6.20$, $\tilde{x} = 2.0$), with a one-sided Welch’s t-test yielding a t-statistic of 92.26 and a p-value $< 0.001$. Similarly, for functions with ad-hoc logs from GitHub Archive ($\mu = 5.62$, $\sigma = 10.21$, $\tilde{x} = 3.0$), the mean complexity exceeded that of all functions ($\mu = 3.64$, $\sigma = 155.06$, $\tilde{x} = 2.0$), with a t-statistic of 7.94 and a p-value of $9.76 \times 10^{-16}$. These results indicate that functions containing ad-hoc logs are significantly more complex than general function samples in both datasets.

\vspace{5pt} 
\noindent\fbox{\begin{minipage}{0.47\textwidth}
\textbf{Finding 1C}: Ad-hoc logs were more frequently placed in the functions with higher cyclomatic complexity. This finding suggests that developers need more assistance in understanding execution behavior in complex logic, and use ad-hoc logs to do so.
\end{minipage}}
\vspace{5pt} 
\subsection{\textbf{RQ2} What do developers put in the log statement?}

\textbf{Ad-hoc Logs with Variable Labels.}
Most logs accept only one (72.6\%) or two (24.2\%) variables or literals. Additionally, echoing the findings from prior work~\cite{jiang_log-it_2023}, developers often add labels to help trace logged variables back to their source code within a stream of console output. Our data suggest that, if a log has two arguments, 84.7\% contain at least one literal, and of those, 51.5\% include the name of the other argument within the literal. For example, \emph{console.log('Results: ', results);} is a case where the developer labels the variable to facilitate locating it in the output console stream.

\begin{figure}
    \centering
    \includegraphics[width=0.9\linewidth]{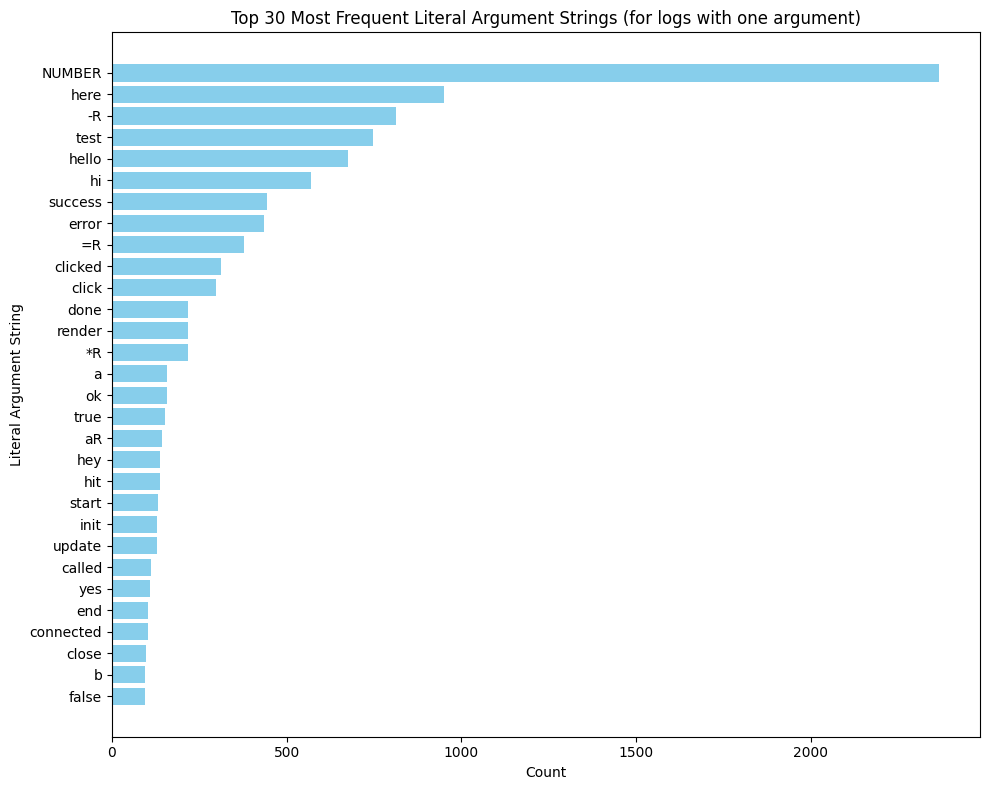}
    \caption{Top 30 Most Frequent Literal Argument (For Logs with One Argument)}
    \label{fig:figure_literals}
    % \vspace{-.2in}
\end{figure}

\textbf{Ad-hoc Logs Formats.}
Figure~\ref{fig:figure_literals} shows the top 30 most frequent literal contents in ad-hoc logs. The symbol \emph{R} represents characters that are repeated more than three times; for instance, \emph{=R} indicates a sequence of repeated equal signs. As shown, the most frequent literal is a number, followed by \emph{here}, and then a sequence of repeated dash signs. Many literals serve as formatting strings or execution order anchors, such as sequences of dashes or equal signs. Developers also use template strings, function names, or brief identifiers (\eg \emph{render}, \emph{hi}, or \emph{hit}), and sometimes formatting functions, to locate or organize outputs. We also analyzed the \emph{CallExpressions} that appear in ad-hoc logs. Among these, the most frequently used function is \emph{JSON.stringify} (14.81\%), a formatting tool that structures complex objects.

In live-stream video coding, the percentages are similar for the log parameter count, where the logs contain one (69.6\%) or two (19.6\%) arguments, and 15 out of 18 (83.3\%) two-argument logs are structured as label string and variables. Furthermore, 9 out of 15 (77.8\%) contain the name of the variable. In addition, similar to the JSON.stringify function, developers using C/C++ tend to create formatting functions of their own. In V4, V7, V12, the developers wrote their own debug function \smalltt{infof}, \smalltt{dbg}, \smalltt{dbgprintf} to simplify forming a more readable log output. When logging variables without native serialization methods, the developers in V4 and V7 wrote helper functions to output variables in a readable string.

\vspace{5pt}
\noindent\fbox{\begin{minipage}{0.47\textwidth}
\textbf{Finding 2}: To locate the variable and to understand execution order on the fly, developers structure the output of ad-hoc logs through labeling and formatted strings akin to findings from prior work~\cite{jiang_log-it_2023}.
\end{minipage}}
\vspace{5pt}

\subsection{\textbf{RQ3}: Why do developers use ad-hoc logs?}
To answer RQ3 and RQ4, we shifted our focus to live-streamed data to address more qualitative questions and gain insight into why and how developers use ad-hoc logs. From analyzing and coding 15 live-stream videos, we summarize our key findings as follows:

\textbf{Understanding Program States}: Developers often use ad-hoc logs to observe or compare program states for several reasons. A common reason is to capture states that are not immediately clear from the source code, such as states introduced by a third-party library or fetched via a web API. For example, the developer in V3 expressed uncertainty about a library function call's effect and logged a boolean value triggered by a GUI click event, noting, ``\textit{Does it return true when option changes, or does it return true when it set it on? That's the thing I don't know. Let's find it out.}'' Developers also log multiple states to cross-reference their values, ensuring they align with expectations. For instance, the developer in V13 checked three boolean values with logs to iteratively verify they matched expectations. In other cases, logs are used to check the state of a single variable or condition, providing quick feedback on specific program values. 
To understand the initial program states when debugging the function, developers also inserted the logs right after the function to inspect the parameters. For example, the developer in V9 log the parameter right after the function to check how the function works, \textit{``alright, let's see what this does.''}
Developers also crafted their own formatting function when dealing with complex or non-readable program states. For example, the developer in V4 working on the curl project introduced a function \texttt{what2name} to transform a constant integer to a string, which increased the readability of logging outputs.%\jim{is ``cURL'' the project name?}

\textbf{Understanding Execution Flow}: Logging is often used to understand or verify code execution, particularly for those in asynchronous functions or executions from third-party libraries. For example, V13 inserted an ad-hoc log with ``{-}{-}{-}{-}{-}'' in a lifecycle method, noting, ``\textit{It gives me a spot to look at... yep, looks like the componentDidUpdate did not get called when I set the props.}'' Another reason is sanity checking, where developers insert strings or variables to confirm that the program reaches specific points. For example, V1 logged “IM GOING TO RUN HERE” to verify if the software actually reach to that specific point.

\vspace{5pt}
\noindent\fbox{\begin{minipage}{0.47\textwidth}
\textbf{Finding 3}:  Developers inserted ad-hoc logs to understand program states and execution flow, especially those are not immediately clear to them such as asynchronous functions or states introduced from third-party libraries. They also used these logs to check their hypotheses of expected runtime behavior and perform sanity checks.
\end{minipage}}
\vspace{5pt}

\subsection{\textbf{RQ4}: How do developers manipulate (insert, revise, and remove) ad-hoc logs to achieve their goal?}
Our study revealed that developers used various methods to create and refine ad-hoc logs, including iterative adjustments, comparisons, diverse timing and strategies for log removal. Key themes include:

\textbf{Iterations}: Developers iteratively refined logs to gain deeper insights into program states and locate the best logging positions or variables. For example, V15 modified the same log three times, adjusting \emph{console.log(this.props)} to \emph{console.log(this.props.navigation.action)} over several edit-run cycles. Similarly, V7 iterated six times to pinpoint the call stack of a double-click event by moving ad-hoc logs back one level with each attempt, remarking, \emph{``I don't remember how all this works, so we are just gonna print to make sense of it.''}

\textbf{Comparison}: Developers used logs to compare program states, often by including multiple states in a single log statement or spreading them across multiple statements. Adding labels, such as variable names, helped track outputs back to their source code. For example, V4 logged the same variable at two different execution points with different labels to check if their values aligned.

Additionally, developers placed ad-hoc logs in multiple locations to compare execution order across different parts of the program, particularly in complex logic or asynchronous code. For instance, V2 inserted two dummy strings in separate locations and observed their execution order in the log output stream as different events occurred.

\textbf{Diverse Timing of Ad-hoc Log Removal}: As expected, the majority of ad-hoc logs are removed once they have served their purpose: 66\% were removed immediately after seeing their output, and further, 87\% were witnessed or known to be removed, eventually. Of the remaining, 8\% are unknown to have been removed, as their videos end before we could witness their fate, and 5\% were seen to be committed to their repository---3 were intentionally committed, as they evolved into production code, and 2 were labeled as possibly accidental commits. 
Some more interesting examples of these include: (1) V1 stored logs in a \textit{git stash} for potential future debugging, and (2) V13 was prompted to remove a log after seeing its output in the terminal.

\textbf{Diverse Strategies for Ad-hoc Log Removal}: Log removal strategies varied. Aside from the most common method of key-in deletion for ad-hoc logs, V1 chose not to commit the file containing the logs, while V9 commented out log 4 to compare different runs and retain it temporarily. Linter tool highlights in V13 reminded developers to remove logs. V13 also used ad-hoc logging in the external experimental environment \texttt{CodeSandbox} to observe outputs in real time without leaving a trace in the main codebase. However, log management was not always deliberate—V15 forgot to delete two ad-hoc logs, which were later committed to the codebase.

\vspace{5pt}
\noindent\fbox{\begin{minipage}{0.47\textwidth}
\textbf{Finding 4}: We observed two main strategies when authoring the ad-hoc logs: (1) Iterative improving the contents or the locations of the ad-hoc logs. (2) Placing multiple logs simultaneously to capture information at once. The varied timing and diverse strategies for removing ad-hoc logs suggest that their removal is just as ad-hoc as their creation.
\end{minipage}}
\vspace{5pt}

%% file: src/discussion.tex
\section{Discussion~\label{section:discussion}}
% \subsection{Implications}
\mysubsection{Perils and Promises of Learning from Mistakes}
As Henry Lieberman noted in 1997~\cite{lieberman_debugging_1997}, ``debugging is the dirty little secret of computer science.'' This ``dirt'' has only accumulated over time, with using debuggers to understand runtime behavior often being regarded as the default solution, leading some developers to feel ambivalent or even reluctant to admit their use of ``printf debugging'' techniques~\cite{beller_dichotomy_2018}. Such hesitation limits the research community’s insight into debugging behaviors, like ad-hoc print statement usage, which play a significant role in shaping final patches or feature development. %\jim{I don't understand what you mean by ``debugging traces.''}
Our research uncovers this ``dirt,'' through mining developers' mistakes to offer the community new possibilities to understand ad-hoc logging practices.

Alaboudi and LaToza's analysis of live-streaming data highlighted that software development is an iterative process involving frequent edit-run cycles~\cite{alaboudi_edit_2021}. However, these small cycles are often hidden within local development environments, limiting the feasibility of large-scale studies. Previous research has addressed this challenge by either conducting costly lab studies~\cite{gilmore_models_1991}, analyzing live-streaming videos~\cite{alaboudi_exploratory_2019}, or requiring developers to install intrusive IDE plugins~\cite{aniche_how_2022, beller_dichotomy_2018}. Our method opens new possibilities for scalable insights into these cycles.

While our collection of ad-hoc logs from revision-control systems enables large-scale analysis, there are limitations: (1) the mined data provides only a snapshot of logs unintentionally committed to version control, making it difficult to observe iterations in the log-authoring process; (2) although the surrounding source code offers some context, it does not capture the developers' rationale for adding, removing, or evaluating the utility of these logs. To gain deeper insight into the context, reasoning, creation, and deletion of these logs, we mined and analyzed the live-streaming data.

Moreover, the distribution of the repository reveals that our dataset, similar to those in prior research~\cite{kalliamvakou_promises_2014}, is skewed toward smaller, personal, and often inactive repositories. We chose not to explicitly exclude personal projects, as this distribution presents a double-edged sword for our analysis. Whereas large, well-maintained open-source projects often enforce linter checks on pull requests to prevent ad-hoc logs~\cite{linter}, personal projects---wherein maintainers might carelessly push ad-hoc logs---allowed us to perform large-scale analysis. Future researchers using our dataset or similar methods should proceed with caution, as data without fine-grained filtering may not fully generalize to expert developers or contributors in collaborative open-source projects.

\mysubsection{Glimpses into the Labyrinth of Complex Software Execution}
Each logging statement placed by developers offers a glimpse into the layered labyrinth of execution. Our analysis captured these fleeting glints to approximate their intentions. Finding 1B reveals that ad-hoc logs were often placed within asynchronous functions, callbacks, and the lifecycle methods of third-party libraries—a sign of developers reaching to grasp execution flows beyond their immediate control. Similarly, Finding 1C shows ad-hoc logs appearing more frequently in functions with higher cyclomatic complexity, suggesting the challenges developers face in unraveling these intricate structures.

While these findings may seem intuitive, the collected glints shed light on re-examining previous metrics and refining our understanding of software complexity. There has long been debate over which metrics—Cyclomatic Complexity, Cognitive Complexity, or even custom models—can best capture the difficulties perceived by developers~\cite{shepperd1988critique, scalabrino_automatically_2017}. However, as discussed before, human studies are costly. Future studies could leverage our empirical dataset to revisit these debates or to evaluate new complexity metrics. Similarly, computer-science educators and software maintainers can observe students’ or developers’ logging behaviors to approximate program complexity, helping them decide when to initiate refactoring or when to adjust the difficulty level.

A single glance is often insufficient to unravel the maze of execution traces. As seen in Finding 4, developers placed logs iteratively, each time drawing closer to understanding the states and execution flows necessary to complete their tasks. These findings suggest that initial logs are often suboptimal for enhancing comprehension. Future research could draw on production log studies to identify optimal logging locations~\cite{fu_where_2014, li_where_2020} and variables~\cite{liu_which_2021}, helping developers reach understanding with fewer steps.

Not only do logging statements need to be placed thoughtfully and contain relevant content; their presentation must also sharpen each glimpse into the call stacks or data structure. Finding 2 shows that many ad-hoc logs include labels and formatting strings, with developers often investing time to create custom debug functions for added clarity. Finding 3 further reveals that to understand or compare states and execution flows, developers carefully craft anchor strings and structure complex objects. Given the rapid pace of program execution relative to human perception, large volumes of log outputs often flood the screen quickly. Consequently, developers must map console outputs back to the source code (spatial mapping) and, in the context of asynchronous functions or third-party libraries, use anchors to mark specific execution timings (temporal mapping). While Log-it has made initial progress in supporting easier mapping~\cite{jiang_log-it_2023}, future tools could further enhance these mappings by incorporating prior works in visualizations~\cite{turcotte_drasync_2022} or recommendation systems~\cite{li_towards_2020}.

\mysubsection{Threats to Validity} 

\mysubsubsection{Internal Threats to Validity} For the mined ad-hoc logs, a key internal threat is that some logs classified as ad-hoc may actually be removed production logs, which could affect the accuracy of our findings. To address this, we manually labeled 100 sample commits and found that only a small portion were not used in an ad-hoc manner. In coding the live-streaming data, researcher bias is a common concern due to the subjective nature of qualitative analysis. We mitigated this by involving a second researcher and holding iterative discussions to minimize uncertainty and resolve any disagreements.

\mysubsubsection{External Threats to Validity} The findings may not generalize to other languages, such as Java and Python. However, we provide scripts to facilitate similar research in different programming contexts. The external validity of the live-stream videos, including representativeness and selection criteria, has been discussed and addressed in the original research by Alaboudi and LaToza~\cite{alaboudi_exploratory_2019}.

%% file: src/related_works.tex
\section{Conclusion~\label{section:conclusion}}
We collected and analyzed 548,880 ad-hoc logs that were accidentally pushed to public version-control systems, along with qualitative analysis from 15 live streaming videos. Our analysis reveals a tendency among developers to insert logs within complex execution flows, such as asynchronous processes, callbacks, and library lifecycles, making them harder to interpret. Additionally, we build upon previous qualitative insights from prior work~\cite{jiang_log-it_2023}, offering empirical evidence that developers often format and label log outputs. We hope this study stimulates further discussion on the challenges of understanding dynamic software behaviors.

In future work, we plan to extend our analyses to include other programming languages and perhaps compare and contrast ad-hoc log usage across these to find trends and patterns. 
Moreover, development-environment tools may be developed to assist with some of the common practices, patterns, and challenges that we observe in our findings here.
\section{Acknowledgment}
We extend our gratitude to the anonymous reviewers for their thorough and insightful feedback. We would also like to thank Heng Du, Junling Wang, Xiaotian Su, Zeyu Xiong, Gustavo Umbelino, Behnood Masoudi, Monil Narang for providing feedback and support.
This research is partially supported by the Dieter Schwarz Stiftung Foundation.